\documentclass[journal]{IEEEtran}
\ifCLASSINFOpdf
\else
\usepackage[dvips]{graphicx}
\fi
\usepackage{url}
\usepackage{graphicx}
\usepackage[colorlinks=true, allcolors=blue]{hyperref}
\usepackage[utf8]{inputenc} 
\usepackage[T1]{fontenc}    
\usepackage{hyperref}       
\usepackage{url}            
\usepackage{booktabs}       
\usepackage{amsfonts}       
\usepackage{nicefrac}       
\usepackage{microtype}      
\usepackage{xcolor}         
\usepackage{amsmath}
\usepackage[square,sort,comma,numbers]{natbib}
\usepackage{algorithm} 
\usepackage{algpseudocode}
\usepackage[noend]{algcompatible}\usepackage{amsmath}
\usepackage{hyperref}
\usepackage{amsmath, mathtools, amsfonts, bm, amssymb, amsthm,graphicx, caption, subcaption,multirow}
\usepackage{algorithm} 
\usepackage{algpseudocode}
\usepackage{bm, bbm}
\usepackage{color}
\usepackage{comment}
\usepackage{adjustbox}
\DeclareMathOperator*{\argmin}{\mathrm{argmin}}
\DeclareMathOperator*{\argmax}{\mathrm{argmax}}

\newcommand{\x}{\bm{x}}

\newcommand{\X}{{\bm{X}}}

\newcommand{\e}{\bm{e}}
\newcommand{\y}{\bm{y}}

\newcommand{\U}{{\bm{U}}}

\newcommand{\R}{{\bm{R}}}
\newcommand{\M}{\bm{M}}

\newcommand{\I}{\bm{I}}
\newcommand{\Y}{\bm{Y}}

\newcommand{\Q}{{\bm{Q}}}

\newcommand{\SE}{\mathrm{SD}}  

\newcommand{\A}{\bm{A}}

\newcommand{\bi}{\begin{itemize}} \newcommand{\ei}{\end{itemize}}
\newcommand{\ben}{\begin{enumerate}} \newcommand{\een}{\end{enumerate}}

\renewcommand{\b}{\bm{b}}

\newcommand{\B}{\bm{B}}

\newcommand{\Bstar}{{\B^\star}}   

\newcommand{\Ustar}{\U^\star{}}
\newcommand{\Xstar}{\X^\star}
\newcommand{\xstar}{\x^\star}

\newcommand{\sigmax}{{\sigma_{\max}^*}}

\newcommand{\sigmaxTwo}{{\sigma_{\max}^{*2}}}

\renewcommand{\P}{\bm{P}}

\newcommand{\bea}{\begin{eqnarray}}
\newcommand{\eea}{\end{eqnarray}}

\newcommand{\kron}{\otimes}

\newcommand{\eps}{\epsilon}

\renewcommand{\b}{\bm{b}}
\renewcommand{\forall}{\text{ for all }}

\newcommand{\lV}{\lVert} \newcommand{\rV}{\rVert}

\newcommand \lv{\|}
\newcommand \rv{\|}

\renewcommand{\U}{{\bm{U}}}

\begin{document}
\title{Locally Permuted Low Rank Column-wise Sensing}
\author{Ahmed Ali Abbasi and Namrata Vaswani, \IEEEmembership{Fellow, IEEE}
	\thanks{The authors are at the Department of Electrical Engineering, Iowa State University. Email: namrata@iastate.edu }}

\maketitle

\begin{abstract}
We precisely formulate, and provide a solution for, the Low Rank Columnwise Sensing (LRCS) problem when some of the observed data is scrambled/permuted/unlabeled.	 This problem, which we refer to as permuted LRCS, lies at the intersection of two distinct topics of recent research: unlabeled sensing and low rank column-wise (matrix) sensing.
We introduce a novel generalization of the recently developed Alternating Gradient Descent and Minimization (AltGDMin) algorithm to solve this problem. We also develop an alternating minimization (AltMin) solution. 
We show, using simulation experiments, that both converge but Permuted-AltGDmin is much faster than Permuted-AltMin.
\end{abstract}

\IEEEpeerreviewmaketitle

\section{Introduction}
In this work, we precisely formulate, and develop a novel solutions for solving, the Low Rank Columnwise Sensing (LRCS) problem \cite{lrpr_gdmin} when some of the observed data is scrambled/permuted/unlabeled, e.g., due to record linkage errors. This permuted LRCS problem can also be understood as the well-studied multi-view unlabeled sensing problem with two modifications: (i) different sensing matrices for each column; and (ii) a low rank assumption on the matrix formed by the unknown signal sequence.  As we explain in Sec. \ref{relwork}, the single and multi-view unlabeled sensing problems have been extensively studied and so has the un-permuted LRCS problem. Potential applications of permuted-LRCS  include (i) multi-task representation learning \cite{lrpr_gdmin,netrapalli} with record linkage errors that cause some rows of the observed data to get permuted, and (ii) LR model based accelerated dynamic MRI  \cite{lrpr_gdmin,lrpr_gdmin_mri} with permutation errors due to k-space trajectory coding mistakes.

\subsection{Problem Setup and Notation}
For a low rank matrix $\Xstar \in \mathbb{R}^{n \times q}$, with $rank(\Xstar) = r \ll \min(n,q)$, we observe $m \ll \min(n,q)$ permuted column-wise measurements $\y_k$. That is,
\begin{equation}
	\y_k := \P^* \A_k \x^*_k  \forall k \in [q], \label{eq:model}
\end{equation}
where $\A_k \in \mathbb{R}^{m \times n}$ is the known measurement matrix,  $\x^*_k$ is the $k$-th column of {\em unknown} $\X^*$, and
$\P^*$ is an $m \times m$ {\em unknown} permutation matrix. 
Given $\y_k$ and $\A_k$,  the goal is to recover $\X^*$ and $\P^*$. To make our problem tractable with $m<n$, we assume  that
\\ (i) the same permutation $\P^*$ acts on all columns of $\Y$ and $\P^*$ is block-diagonal with blocks of size $s$ with $s$ small enough so that $m/s > r$  ($s$-local permutation);
and
\\
ii) right singular vectors of $\X^*$ are incoherent, that is  $\lV \x_k^* \rV_2 \leq \mu \sigmax \sqrt{r/q} \forall k \in [q]$, where $\sigmax$ denotes the largest singular value of $\X^*$.
\\
Both of these assumptions are standard ones commonly used in the multi-view unlabeled sensing or LRCS literature respectively. See Sec. \ref{relwork}.

We factor $\Xstar := \Ustar \Bstar$ where $\Ustar$ is the $n \times r$ matrix of left singular vectors of $\Xstar$ with nonzero singular values.

Formally, the set of $m \times m$ $s$-local permutation matrices  $\Pi_{m,s}$ is defined as
\begin{align}
\Pi_{m,s} \coloneq &\{\P^* \mid  \P^* = \text{blockdiag} (\P_1^*, \cdots, \P_{m/s}^*), \P_{i}^* \in \Pi_{s} \} \label{eq:sLcl}
\end{align}	
where $\Pi_s$ is the set of all $s \times s$ permutation matrices, i.e., 
\[
\Pi_s \coloneq \{ \P \mid \P \in \{0,1\}^{s \times s}, \P\mathbbm{1}_s = \mathbbm{1}_{s}, \P^\intercal\mathbbm{1}_s = \mathbbm{1}_s \}
\]
and $\mathbbm{1}_s$ denotes the all-ones vector of dimension $s$.



\subsubsection{Notation} 
To keep notation simple, we assume that $m/s$ is an integer. Bold capitalized letters, e.g. $\Y$, denote matrices, bold small letters, e.g., $\y_k$, denote vectors, and un-bolded small letters denote scalars. We use $||\M||_2$, or just $||\M||$, to denote the (induced) 2-norm and $||\M||_F$ to denote the Frobenius norm of a matrix $\M$.
We use $\M^{\dagger} \triangleq (\M^\intercal\M)^{-1}\M^\intercal$ to denote the pseudo-inverse of a tall matrix $\M$. $\A_{k,i} \in \mathbb{R}^{s \times n}$ denotes the sub-matrix formed by the rows in the $i$-th block of $\A_k \in \mathbb{R}^{m \times n}$; thus $\A_k^\intercal = [\A_{k,1} \mid \cdots \mid \A_{k,m/s}]$. 
$\mathrm{QR}(\M)$ maps $\M$ to $\Q$ such that $\M = \Q \R$ is the $\mathrm{QR}$ decomposition of $\M$; we restrict to tall $\M$, i.e., $\M$ with more rows than columns.
For two $n \times r$ matrices with orthonormal columns, we use $SD(\U,\Ustar):=||(\I - \Ustar \Ustar^\top)\U||$ to denote the subspace distances between their column spans.

\begin{figure}
	\begin{center}
		\includegraphics[width = 0.40 \linewidth]{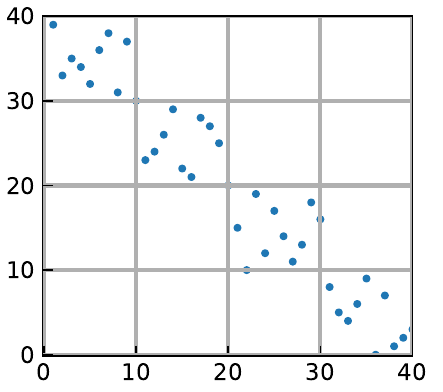}
\vspace{-0.07in}
		\caption{An example of the $s$-local permutation model with $4$ blocks of size $s = 10$ each.}
\vspace{-0.25in}
		\label{fig:permModels}
	\end{center}
\end{figure}

\subsection{Relevant Literature} \label{relwork}
\subsubsection{Single View Unlabeled Sensing (ULS)}
\label{subsec:ulsRlvntWork}
In single-view unlabeled sensing (ULS), given scrambled observations $\y = \P^* \A \x^*$, and sensing matrix $\A \in \mathbb{R}^{m \times n}$, the problem is to recover the \textbf{vector} $\x^* \in \mathbb{R}^n$. Note that the permutation matrix $\P^*$ is unknown, which makes the problem different and more challenging compared to the standard linear inverse problem.   The work in \cite{unnikrishnan2018unlabeled} formulated the single-view ULS problem and established that $m = 2n$ measurements are both necessary and sufficient to recover $\x^*$.  For single-view ULS, algorithms based on branch and bound and expectation-maximization (EM) are proposed in \cite{emiya, concave, header_free}, which are suitable for small problem sizes. A modified EM approach is proposed in \cite{abid2018stochastic}.

\subsubsection{Multi View ULS}
The problem is to recover $\X^* \in \mathbb{R}^{n \times q}$ from permuted matrix measurements $\Y := \P^*\A\X^*$. In this problem, $\A$ is the same for all columns of $\Xstar$ making it very different from the LRCS problem. (i) This needs $m>n$, even if the permutation were known. (ii) Also, it needs full rank $\X^*$ and large enough $q$ to have enough diversity to make the permutation recovery possible \cite{Levsort}.
We also do not assume low rank on $\X^*$. 
Several works also assume either a partially shuffled model \cite{slawski-single,slawski2020sparse,slawski_two_stage}  or a block-diagonal model \cite{abbasi2021r, abbasi2022r} for $\P^*$.  The works \cite{slawski2020sparse, snr, slawski_two_stage, icml, emiya} propose methods based on sparse subspace clustering, bi-convex optimization, robust regression, and spectral initialization,  and branch-and-bound optimization, respectively. Algorithms based on graph matching and alternating minimization with a suitable initialization were proposed in \cite{abbasi2021r, abbasi2025alternating}, respectively.  More recently, ULS with sparse $\x^*$ was studied in \cite{peng2021homomorphic, akrout2025unlabeled}.

\subsubsection{The LRCS problem and AltMin and AltGDmin algorithms} The LRCS problem, or its generalization called the LR phase retrieval problem, have been extensively studied in the last five years \cite{lrpr_tsp,lrpr_icml, lee2019neurips,lrpr_best,lrpr_gdmin,netrapalli_mtrl}. Well known solutions include a very slow convex relaxation \cite{lee2019neurips}, a faster alternating minimization (AltMin) algorithm \cite{lrpr_icml,lrpr_best}, and an even faster gradient descent (GD) based solution approach called Alternating GD and minimization (AltGDmin) \cite{lrpr_gdmin}. All results assume right singular vector incoherence or a stronger version of it.
Both AltMin and AltGDmin factor the unknown LR matrix $\X$ as $\X = \U \B$ with $\U$ being $n \times r$ and $\B$ being $r \times q$. AltGDmin is a novel modification of the AltMin approach for problems such as LRCS in which minimization w.r.t. one of the two variable subsets, $\B$, is much faster than that w.r.t. to the other, $\U$. The reason is that the latter decouples column-wise.
After initializing $\U$ using a carefully designed spectral initialization, it alternates between updating $\B$ (keeping $\U$ fixed) using minimization, and updating $\U$ (keeping $\B$ fixed)  using one GD step. The updated $\U$ is orthonormalized using QR decomposition. 
In more recent work \cite{lrmc_altgdmin}, AltGDmin has been studied for communication-efficient LR matrix completion.

\subsubsection{Other tangentially related work} Other somewhat related work includes 
\cite{yao2021unlabeled, yao2024unlabeled, mruc,vogelstein2015fast}.

\subsection{Our Contribution}
We introduce and precisely formulate the Permuted LRCS problem and the required locally permuted assumption. We develop a novel generalization of the AltGDmin algorithm to handle the permutations.
Considering squared loss, our goal is to minimize the objective function  $f(\P,\U,\B) \coloneq  \sum_{k=1}^{k=q} \lV \y_k - \P \A_k \U \b_k \rV_2^2$ under the constraints that $\P$ is an $s$-local permutation matrix and $\U$ has orthonormal columns.
That is, we need to solve
\begin{equation}
	\min_{\U \in \mathbb{R}^{n \times r}\mid \U^\intercal \U = \mathbf{I}, \B \in \mathbb{R}^{r \times q} , \P \in {\Pi}_{m,s}} f(\P,\U,\B),
	\label{eq:obj}
\end{equation}
where $\Pi_{m,s}$ is defined in \eqref{eq:sLcl}.

The AltGDmin algorithm has so far been used only for problems in which the set of unknowns are split into two variable subsets $\U$ and $\B$. However, in our current problem, the natural split-up involves three subsets $\P, \U, \B$. We develop a generalization of AltGDmin that updates the three subsets in sequence with using minimization for updating $\P$ and $\B$, and GD for updating $\U$. This is done because the minimizations over $\P$ and over $\B$ are much faster than that over $\U$.
Keeping $\P$ and $\U$ fixed, the update of $\B$ involves solving column-wise least squares (LS) problems with total complexity only order $mqnr$.
Keeping $\U,\B$ fixed, we show below that, minimizing for $\P$ is the well-known linear assignment problem (LAP) which can be solved exactly using the Hungarian assignment algorithm \cite{munkres1957algorithms}. This step has complexity of order $mqn$. The update of $\U$ by a full minimization has a complexity of order $mq (nr)^2$. Hence, AltGDmin uses GD for updating $\U$ with a cost of order $mqnr$.

A second challenging aspect of our problem is that there is no easy way to initialize the permutation matrix. We instead develop a novel modification of the LRCS initialization by adapting the collapsed initialization idea introduced for the ULS problem in \cite{abbasi2021r}.

We also develop an AltMin based solution.  We show, using simulation experiments, that both converge but Permuted-AltGDmin is much faster.



\section{Proposed Algorithm: Permuted AltGDMin}



As we explain below, the cost of exactly minimizing over $\P$, keeping $\B$ and $\U$ fixed, in \eqref{eq:obj} is $O(ms^2 + nq(m + r) + msq) = O(mqn)$.  That for  $\B$, keeping $\P$ and $\U$ fixed, is $O(mr^2q + mnrq) = O(mqnr)$. In contrast, exact minimization over $\U$, keeping $\B$ and $\P$ fixed,  costs $O(mq n^2r^2)$. This is much higher due to quadratic dependence on $n$. As such, we update $\U$ using a single GD iteration followed by an $n \times r$ QR decomposition at a lower cost of $O(mqnr + nr^2) = O(mqnr)$.


\begin{algorithm}[t]
	\caption{Perm-AltGDMin}
	\begin{algorithmic}[1]
		\label{algorithm}
		\REQUIRE  Observations $\Y$,  measurement matrices $\A_k $,  rank $r$,  step size $\eta$, number of iterations $T$, block size $s$.
		\FOR {$k \in \{1,\cdots,q\}$}
			\STATE{Form  $\A_{k, cllps}$ and $\y_{k,cllps}$ according to \eqref{eq:cllps}}\vspace{0.15em}
		\ENDFOR
		\STATE{$\M^{(0)} \gets \sum_{k=1}^{k=q}\A_{k,cllps}^\intercal \y_{k,cllps} \e_k^\intercal$}\vspace{0.15em}
		\STATE{$\U^{(0)} \gets  \text{top $r$ left-singular vectors of} \M^{(0)}$}\vspace{0.1em}
		\FOR {$k \in \{1,\cdots,q\}$}
		\STATE{$\b_k^{(0)} \gets ( \A_{k,cllps}\U^{(0)})^{\dagger} \y_{k,cllps}$,  see \eqref{eq:LSCllps}}
		\STATE{$\widehat\y_k^{(1)} \gets \A_k \U^{(0)}\b_k^{(0)}  $}	
		\ENDFOR
		\FOR {$t \in \{1,  \cdots,T\}}$ 
		\FOR {$i \in \{1, \cdots, m/s\}$}
		\STATE{$\P^{(t)}_i = \argmax_{\P_i \in \Pi_{s}} \langle \Y_i ,\P_i \widehat \Y_i^{(t)}  \rangle$, see \eqref{eq:samePermBlcks}}
		\ENDFOR
	\STATE{$\P^{(t)} \gets \text{blkdiag}(\P_1^{(t)}, \cdots, \P_{m/s}^{(t)})$ \vspace{0.35em}}
	\STATE{$\nabla_{{\U}} f \gets \sum_{k=1}^{k=q} (\P^{(t)} \A_k)^{\intercal}(\P^{(t)}\widehat\y_k^{(t)}  - \y_k)\b_k^{(t)\intercal}$	}	\vspace{0.35em}
	\STATE{$\U^{(t)} \gets \mathrm{QR}(\U^{(t-1)} - \eta \nabla_\U f)$, see \eqref{eq:UGradUpdt}, \eqref{eq:gradExprsn}}  \vspace{0.31em}
	\FOR {$k \in \{1,\cdots,q\}$}
	\vspace{0.15em}
	\STATE{$\b_k^{(t)} \gets (\P^{(t)} \A_k\U^{(t)})^{\dagger} \y_k$, }	\vspace{0.15em}
\STATE{$\widehat\y_k^{(t+1)} \gets \P^{(t)}\A_k \U^{(t)}\b_k^{(t)}  $}
\ENDFOR

\ENDFOR
	
	\STATE { \bfseries {Return} $\U^{(T)}, \B^{(T)}$ }
\end{algorithmic}

\label{algo}
\end{algorithm}

\subsection{Perm-AltGDMin Initialization $(t=0)$}
\label{subsec:cllpsInit}

\subsubsection{Formation of collapsed system}
Since there is no good way to initialize the permutation matrix, we instead develop a modification of the collapsed initialization idea that was introduced in \cite{abbasi2021r} for unlabeled sensing. Because of the $s$-locality assumption on the permutation matrix, if we sum consecutive sets of $s$ measurements, we would eliminate the permutation matrix. To be precise, $\mathbbm{1}_s^\intercal \P^*_i = \mathbbm{1}_s^\intercal$ since $\P^*_i$ is an $s \times s$ permutation matrix. Define a $(m/s) \times m$ matrix that is block diagonal with $m/s$ blocks (each block is a row of length $s$),
\[
\bm{C}_{cllps}: = blkdiag(\mathbbm{1}_s^\intercal, \mathbbm{1}_s^\intercal, \dots, \mathbbm{1}_s^\intercal)
\]
For each $k \in [q]$, define
\begin{align}
\y_{k,cllps}:= \bm{C}_{cllps} \y_k, \ \A_{k,cllps}:=  \bm{C}_{cllps} \A_k
\label{eq:cllps}
\end{align}
Thus, $\A_{k,cllps} \in \mathbb{R}^{(m/s) \times n}$ and $\y_{k,cllps} \in \mathbb{R}^{(m/s) \times 1}$ and these satisfy
\[
\y_{k,cllps} = \A_{k,cllps} \xstar_k
\]
\subsubsection{$U^{(0)}$ initialization (line 4 of Algorithm \ref{algo})}
$\U^{(0)}$ is initialized by the top $r$ left-singular vectors of the matrix
\[
\sum_{k=1}^q \A_{k,cllps}^\intercal \y_{k,cllps} \e_k^\intercal
\]
This is the same initialization as in \cite{lrpr_gdmin}, but without the truncation step, which was introduced there for theoretical analysis.
\subsubsection{$B^{(0)}$ initialization (line 6)}
Given $\U^{(0)}$, we obtain $\b_k^{(0)} \in \mathbb{R}^{r}$  by minimizing the collapsed least squares problem with \textbf{$(\bm{m/s})$}  measurements and $\A_{k,cllps} \U^{(0)} \in \mathbb{R}^{(m/s) \times r}$ as the sensing matrix, $\b_k^{(0)} = \argmin_{\b} \lV \y_{k,cllps} - \A_{k,cllps}\U^{(0)} \b \rV_2^2$. This has the closed form solution:
\begin{equation}
	 \b_k^{(0)} =  (\A_{k,cllps}\U^{(0)})^\dagger \y_{k,cllps}\, \forall k \in [q]. \label{eq:LSCllps}
\end{equation}
Then, $\widehat\y_k^{(1)} \in \mathbb{R}^{m}$ is formed as $\widehat\y_k^{(1)} = \A_k \U^{(0)} \b_k^{(0)}$. 

\subsection{Perm-AltGDMin Iterations ($t \geq 1$)}
Given $\U^{(t)}, \B^{(t)}$, we can obtain $\widehat\y_k^{(t+1)} = \A_k \U^{(t)}\b_k^{(t)} \, \forall k \in [q]$. We use this to first estimate the permutation matrix, followed by then updating $\U$ and $\B$.

An alternate approach can be to also use collapsed measurements for the AltGDmin iterations; in this case the algorithm to use would be exactly the same as that for basic LRCS. As we demonstrate in the simulations section, this is much worse than our approach. The reason is, given a good initial estimate of $\U, \b$, we can get a good estimate of $\P$, and with this, we are able to use many more measurements.



\subsubsection{$\P$-update (line 10 of Algorithm \ref{algo})}
For $t \geq 1$, the permutation matrix $\P^{(t)}$-update is by the following linear assignment problem (LAP)
\begin{equation}
	\P^{(t)} = \argmin_{P \in \Pi_{m,s}} \lV  \Y - \P \widehat\Y^{(t)} \rV_F^2.\label{eq:samePerm}
\end{equation}
To see that \eqref{eq:samePerm}  is an LAP,
\begin{align}
		\argmin_{\P \in \Pi_{m,s}} \lV  \Y - \P \widehat \Y^{(t)} \rV_F^2
		= \argmax_{\P \in \Pi_{m,s}} \langle  \Y, \P \widehat \Y^{(t)} \rangle \label{eq:lapObj},
\end{align}
\eqref{eq:lapObj} follows from noting that for any permutation matrix $\P$, $\lV \P \Y \rV_F = \lV \Y \rV_F$ so that $\argmin_{\P \in \Pi_{m,s}} \lV  \Y - \P \widehat \Y^{(t)} \rV_F^2 = $
$$ \hspace{-0.21em} \argmin_{\P \in \Pi_{m,s}} \lV \Y \rV_F^2 + \lV  \P \Y \rV_F^2 - 2 \langle  \Y, \P\widehat \Y^{(t)} \rangle = \argmax_{\P \in \Pi_{m,s}} \langle  \Y, \P \widehat \Y^{(t)} \rangle.$$
   The objective function in \eqref{eq:lapObj} is a linear function of the optimization variable $\P$, making this a linear assignment problem.  Because $\P^* \in \Pi_{m,s}$ is block-diagonal,  \eqref{eq:lapObj}  simplifies to decoupled updates of smaller sizes.  That is, for  $\P^* = \text{blkdiag}(\P_1^*, \cdots, \P_{m/s}^*)$,   with each block of size $s$,
\begin{align}
	\P^{(t)}_i  &= \argmax_{\P_i \in \Pi_{s}} \langle \Y_i ,\P_i \widehat \Y_i^{(t)}  \rangle \notag \\&= \text{trace}( \widehat\Y_i^{(t)}\Y^{\intercal}_i \P_i) \forall i \in [m/s].\label{eq:samePermBlcks}
\end{align}
Each of the $s$-dimensional $m/s$ LAPs in \eqref{eq:samePermBlcks} can be solved exactly by the Hungarian assignment algorithm in $O(s^3)$ time \cite{munkres1957algorithms}. 
Additionally, the cost of forming  $\widehat\Y$ is $O(nq(m + r))$, followed by the $O( (m/s) \cdot s^2 q) = O(msq)$ cost of forming the $s \times s$ block matrices $\widehat\Y_i \Y_i^\intercal \forall i \in [m/s]$ .  Consequently, the total cost of  updating the $m/s$ blocks in \eqref{eq:samePermBlcks} is $O(ms^2 + nq(m + r) + msq)  = O(mqn)$.

We emphasize here that, while the linear assignment problem \eqref{eq:lapObj} can be solved exactly to find the minimum value and {\em a} minimizer, the minimizer may not be unique. In particular, this means that, in general, there is no guarantee on the quality of the obtained estimate $\P$. We expect the quality of the estimate to depend on how large $q$ is and how close $\widehat\Y$ is to the unpermuted version of the $\Y$, i.e., to $\P^*{}^\intercal\Y$.  We will postpone the analysis of this step, and of the complete approach, to future work.

\subsubsection{$U$-update (lines 12-13 of Algo. \ref{algo})} We update $\U$ by a single gradient descent step, followed by a QR mapping. For $t \geq 1$,
\begin{equation}
\U^{(t)} \gets \mathrm{QR}(\U^{(t-1)} - \eta \nabla_\U f), \label{eq:UGradUpdt}
\end{equation}
where the expression for the gradient is
\begin{equation}
\nabla_{{\U}} f = \sum_{k=1}^{k=q} (\P^{(t)} \A_k)^{\intercal}(\P^{(t)}\widehat\y_k^{(t)}  - \y_k)\b_k^{(t)\intercal}.	\label{eq:gradExprsn}
\end{equation}
We discuss step-size $\eta$ selection in Section \ref{sec:rslts}. The QR step in \eqref{eq:UGradUpdt} ensures that the norms of the iterates $\U^{(t)}$ and $\B^{(t)}$ remain bounded, as discussed earlier. 

\subsubsection{$B$-update (line 15 of Algorithm \ref{algo})}
For $k \in [q]$ and $t \geq 1$,  $\b_k^{(t)}$ is updated by solving the $(m/s) \times r$ least-squares problem $\b_k^{(t)} = \argmin_{b} \lv \y_k - \P^{(t)} \A_k  \U^{(t)} \b \rv_2^2$, with closed form solution $\b_k^{(t)} = (\P^{(t)} \A_k\U^{(t)})^{\dagger} \y_k$.  
Subsequently, $\y_k^{(t+1)} = (\P^{(t)} \A_k\U^{(t)}) \b_k^{(t)}$, and we repeat the $\P$, $\U$ and $\B$ updates outlined above for $T$ iterations.

\subsection{Perm-AltMin algorithm}
AltMin is the same as above except the update of $\U$ involves solving the LS problem keeping $\P, \B$ fixed at previous values. This is much slower since the problem dimension is $mq \times nr$. 

\begin{figure}[t!]
	\begin{subfigure}{0.49 \linewidth}
		\includegraphics[width=1.01 \linewidth]{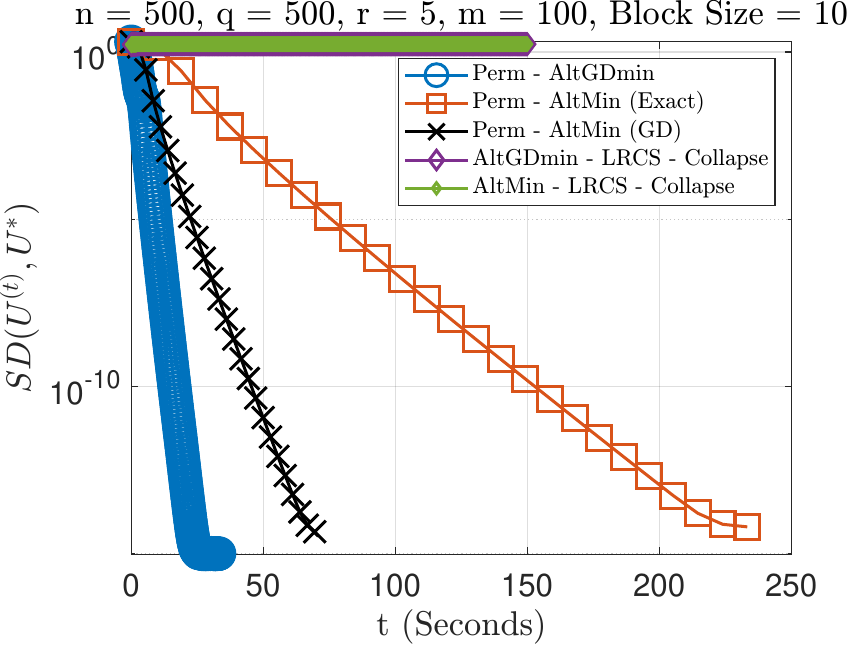}
\caption{\small\sl{Error vs time-taken}}
	\end{subfigure}		
	\begin{subfigure}{0.50 \linewidth}
		\includegraphics[width=1.05\linewidth]{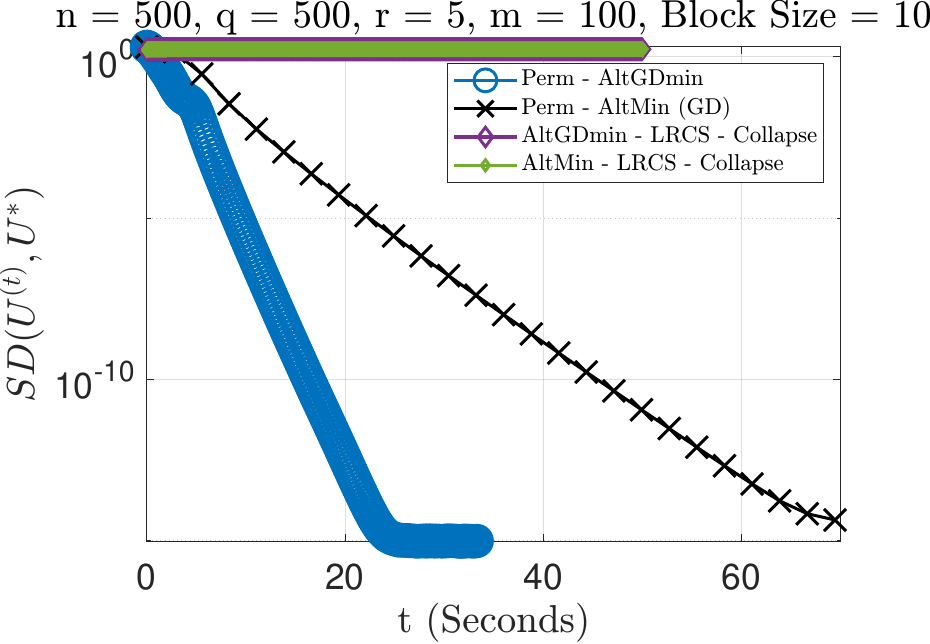}
\caption{\small\sl{Higher resolution of (a)}}
	\end{subfigure}
	\caption{\small\sl{Run-time comparisons.}} 
	\label{fig:runTime}
\vspace{-0.1in}
\end{figure}

\begin{figure}
	\begin{subfigure}{0.49\linewidth}
		\includegraphics[width=0.99\linewidth]{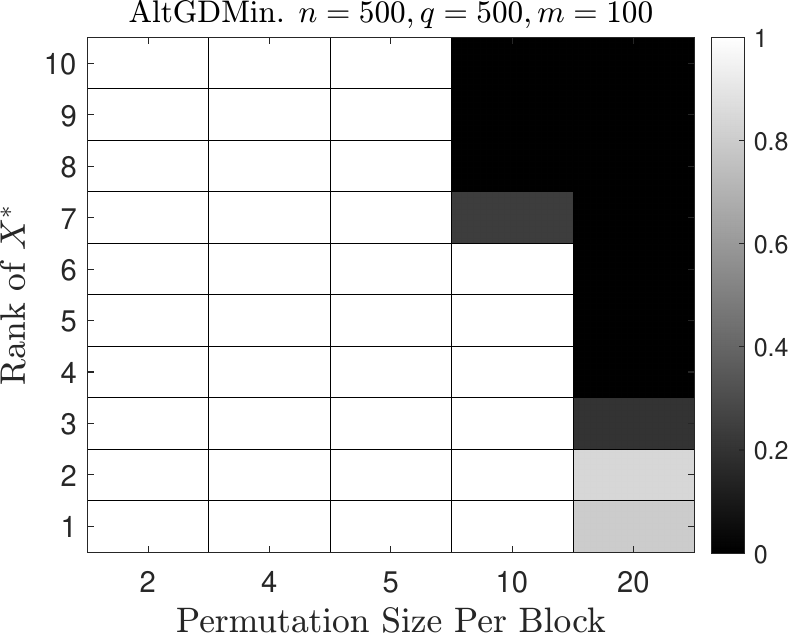}
			\vspace{-0.95em}
\caption{\small\sl{Perm-AltGDmin}}
	\end{subfigure}
	\begin{subfigure}{0.49\linewidth}
		\includegraphics[width=0.99\linewidth]{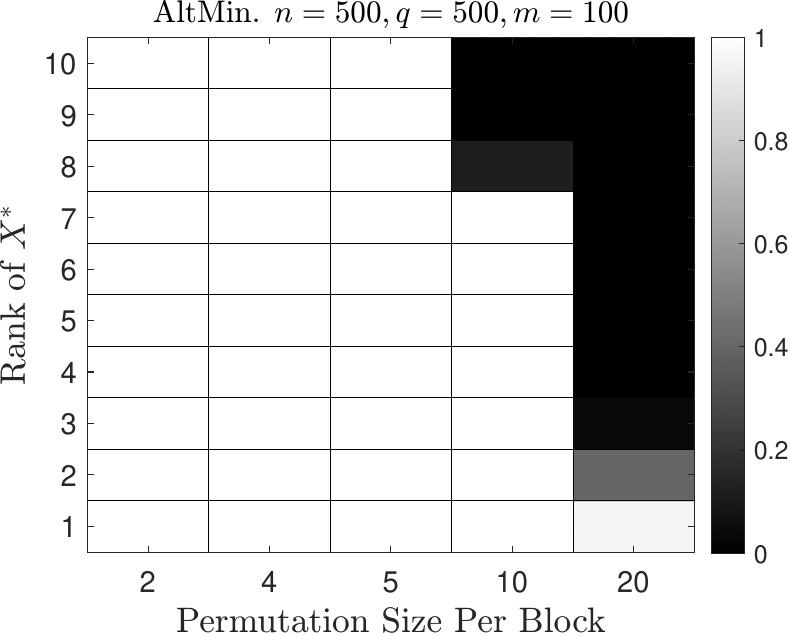}
\vspace{-0.95em}
		\caption{\small\sl{Perm-AltMin}}
	\end{subfigure}
		\caption{\small\sl{Phase transition plots. $\Pr[\SE(\U^{(T)},\U^*) \leq 10^{-10}]$ is plotted against the block permutation size, $s$, and rank of $\X^*$, $r$.}}
	\label{fig:phasetrans}
\vspace{-0.2in}
\end{figure}

\section{Simulation Results} \label{sec:rslts}

\subsubsection{Benchmark methods' description}
We compare our algorithm to two implementations of Alternating Minimization (AltMin).    AltMin alternates exact minimization (least-squares problems) for both $\U$ and $\B$ variables in the objective function \eqref{eq:obj}.   To modify AltMin for the permuted setup, we make two changes: i) add the proposed  $\P$-update \eqref{eq:samePermBlcks} and ii) initialize by the collapsed initialization \eqref{eq:cllps}.  Consequently, both algorithms have the same  initialization $(\U^{(0)}, \B^{(0)})$ (lines 4 and 6 of Algorithm 1)  , $\B$ least-squares (line 15), $\P$ linear assignment update (line 10), and differ only in the $\U$ update (line 13). Compared to $1$ gradient descent iteration  for $\U^{(t)}$ in  AltGDMin followed by a QR step, AltMin updates $\U^{(t)}$ by exact least-squares without the QR step. We also compare with AltGDmin-LRCS-Cllps and AltMin-LRCS-Cllps, which are the  AltGDMin and AltMin method run on the collapsed measurements, without any $\P$-update. These collapsed-measurements-only methods do not converge because the number of collapsed measurements $m/s$ is small compared to the to the higher but permuted number of measurements, $m$.

\subsubsection{Implementation Details} We set the AltGDMin step-size for gradient descent $\U$ update as $\eta = 0.3/(m \sigmaxTwo)$, and estimate $\sigmax \simeq \sigma_{\max}(\widehat \X^{(1)})$. This choice of step-size is suggested for AltGDMin (without permuted measurements) in \cite{lrpr_gdmin}. For the least-squares update of $\U$ in AltMin, we set the step-size as $\eta = 1/L$, where $L = \sum_{k=1}^{k=q}\sigma_{\max}^2(\A_k)\lV \b_k \rV_2^2 $ is the Lipschitz constant of $f(\U) = \sum_{k=1}^{k=q} \lV \y_k  - \A_k \U \b_{k}\rV_2^2$. We compute $L$ only at $t=0$ because computing  it at every iteration, requires reevaluating $\lV \B^{(t)} \rV_F^2$, which is  computationally expensive. We use the backslash operator in MATLAB to solve the least-squares problems. For the linear assignment problem, we use the MATLAB `matchpairs' command.   For a fast implementation, we do not  construct the square matrix $\P^{(t)}$,  instead representing $\P^{(t)}$ by a vector with shuffled entries in the integer range $[1,m]$.

\subsubsection{Synthetic Data Generation} We form rank-$r$ $\X^* = \U^*\B^*$ by generating the left-singular vectors $\U^* \in \mathbb{R}^{n \times r}$ as the othonormal basis of an $n \times r$ Gaussian $\sim \mathcal{N}(0,1)$ random matrix and $\B^* \in \mathbb{R}^{r \times q}$ as a Gaussian matrix.  $\A_k \in \mathbb{R}^{r \times q}$ are also Gaussian random matrices.  $\P^*$ is an $r$-local permutation matrix, that is, an  $m \times m$ matrix, with $m/s$ block-diagonal permutations of size $s$ each.  At each Monte-Carlo run, we only change the permutation matrix $\P^*$, keeping $\A_k$, $\B^*$ and $\U^*$ the same.
\subsubsection{Fig. \ref{fig:runTime} Observations} The run-time plots show that AltGDMin is the fastest algorithm to converge.
The computational complexity of the AltGDMin $\U$-update, which is the complexity of one gradient computation  $O(mnqr)$ and one QR decomposition $O(nr^2)$, is $O(mnqr^2)$. The computation cost of AltMin $\U$-update (exact least squares (LS) using gradient descent) is  $O(\kappa mnqr \log(1/\eps))$, where $\kappa \log(1/\eps)$ is the iteration complexity and $\kappa$  is the strong convexity constant (or condition number) of the  $\U$-update least squares objective function $f(\U) = \sum_{k=1}^{k=q} \lV \y_k  - \A_k \U \b_{k}\rV_2^2$. For $\kappa \log(1/\eps) > r$, the latter cost of exact (LS) minimization (i.e. small $\eps$) is higher.   Also, computing the gradient several times is slow because each computation requires a `for' loop over $q$ terms \eqref{eq:gradExprsn}.
AltMin (Exact) is much slower than AltMin (GD) because it solves a least-squares problem  $[ \b_1^\intercal \kron \A_{1} \mid  \cdots \mid \b_q^\intercal \kron \A_{q}]^{\dagger} \y_{all}$  of dimension $nr$ using matrix-inversion at each iteration with computational complexity $O(mqn^2r^2)$, whereas the latter does not use matrix inversion, instead using several iterations of gradient descent.
\subsubsection{Fig. \ref{fig:phasetrans} Observations} We plot the probability of recovery $\Pr[\SE(\U^{(T)},\U^*) \leq 10^{-10}]$ against the permutation block size and the rank of $\X^*$, for both AltGDMin and AltMin.  As expected, the probability of recovery increases with decreasing rank and decreasing block size $s$, where $m/s$ is the number of blocks in $m \times m $ permutation $\P^*$. For $s$-local $\P^*$, the number of blocks is $m/s$. Therefore, a smaller block size $s$ not only translates to a smaller permutation problem, but also an improved initialization with higher $m/s$ measurements in the collapsed system \eqref{eq:cllps}. A lower value of rank $r$ requires fewer measurements because the number of unknowns in $\U$ and $\B$ are $(n + q)r$. For AltMin, we observe slightly better performance with succesful recovery at block size $s = 10$ and rank $r = 7$, possibly because AltMin fully minimizes $\U$ at every iteration, whereas AltGDMin only does a single gradient descent update.  However, as the results in Fig. \ref{fig:runTime} show, full minimization is computationally expensive and makes the overall algorithm slower.

\section{Conclusion}
We introduced a novel solution approach, called Perm-AltGDMin, for solving the permuted LRCS problem.
Open questions for future work include: (i) analyzing Perm-AltGDmin for this problem, and (ii) studying if a similar approach can be developed for LR matrix completion by modifying the recently studied AltGDmin algorithm for it \cite{lrmc_altgdmin}.


\clearpage
\bibliographystyle{IEEEtran}
\bibliography{tipnewpfmt_kfcsfullpap,uls}

\end{document}